\documentclass[prb,aps,showpacs,twocolumn]{revtex4}
\usepackage{amsmath,amsfonts}

\usepackage{bm}

\usepackage{graphicx}
\usepackage{subfigure}

\usepackage[squaren]{SIunits}
\usepackage{nicefrac}

\usepackage{hyperref}

\DeclareMathOperator{\arctanh}{arctanh}
\DeclareMathOperator{\sgn}{sign}
\newcommand{\signum}[1]{ \sgn \left( #1 \right) }

\newcommand*{\pauli}{ \vec{\sigma} }

\newcommand{\form}[1]{ \,\mathrm{d} #1 \;}
\renewcommand{\vec}[1]{ \bm{ #1 }}
\newcommand{\abs}[1]{ \left| #1 \right| }

\newcommand*{\meV}{\milli\electronvolt}
\newcommand*{\nm}{\nano\meter}

\begin{document}

\title{Spin transport in proximity induced ferromagnetic graphene}

\author{H\aa vard Haugen}
\email{havard.haugen@ntnu.no}
\author{Daniel Huertas--Hernando}
\author{Arne Brataas}
\affiliation{Department of Physics, 
  Norwegian University of Science and Technology,
  N-7491 Trondheim, Norway}

\date{\today}

\begin{abstract}
  Ferromagnetic insulators deposited on graphene can induce
  ferromagnetic correlations in graphene. We estimate that induced
  exchange splittings $\Delta \sim \unit{5}{\meV}$ can be achieved by
  \text{e.g.} using the magnetic insulator EuO. We study the effect of
  the induced spin splittings on the graphene transport properties.
  The exchange splittings in proximity induced ferromagnetic graphene
  can be determined from the transmission resonances in the linear
  response conductance or, independently, by magnetoresistance
  measurements in a spin-valve device. The spin polarization of the
  current near the Dirac point increases with the length of the
  barrier, so that long systems are required to determine $\Delta$
  experimentally.
\end{abstract}

\maketitle

\section{Introduction}

The two dimensional honeycomb lattice of graphene is a conceptual
basis for describing carbon structures such as fullerenes, carbon
nanotubes and individual layers of graphite. The fabrication of free
and stable mono-layers of graphene a few years ago transformed this
concept into an experimental reality that has attracted a tremendous
interest from the research
community.\cite{Novoselov2004sci306,Novoselov2005n438,Zhang2005n438}
The low energy excitations of charge carriers in graphene are similar
to massless relativistic Dirac (or rather Weyl) particles. The
Hamiltonian is~%
\cite{Gonzalez1993,Nomura:cond-mat0606589}
\begin{equation}
 \label{eq:graphene-low-energy-hamiltonian}
 H = -i \hbar v \pauli \cdot \nabla + U(\vec{r}),
\end{equation}
where the velocity $v\approx \unit{10^{6}}{\metre\per\second}$ is the
analogue of the Dirac electrons speed of light (in the sense of
limiting velocity) in graphene and $\pauli = (\sigma_{x}, \sigma_{y})$
is a two dimensional vector of Pauli matrices acting on 2-spinor
states related to the two triangular sub-lattices constituting
graphene's honeycomb lattice.  The approximate
Hamiltonian~\eqref{eq:graphene-low-energy-hamiltonian} is valid near
the Dirac points $K$ and $K^{\prime }$ in the reciprocal lattice. The
two inequivalent Dirac points introduce a two-fold valley
degeneracy.\cite{morpurgo:prl:v97:p196804:y2006}

The carrier concentrations are typically in the range
$\unit{10^{11}-10^{12}}{\centi\metre^{-2}}$, corresponding to a Fermi
wavelength of $\lambda _{F}\approx
\unit{50-100}{\nm}$.\cite{Katsnelson2006nphys,Zhang2005n438} The mean
free path has been reported to be of the order $l_\text{mfp} \approx
\unit{0.4}{\micro\metre}$.\cite{Novoselov2004sci306} With improved
control over the fabrication process of graphene, we expect to see the
realization of even cleaner samples with longer mean free paths.

Spintronics aim to inject, manipulate, and detect spins in electronic
devices. Electrical spin injection in normal metals is routinely
achieved by contacting ferromagnets like Fe, Ni, and Co with normal
metals such as Cu and Al and driving a current through the system. In
semiconductors, electrical spin injection is more challenging because
of the resistance mismatch between the semiconductor and possible
ferromagnetic metal contacts.\cite{Schmidt:prb62} Nevertheless,
spin-injection into a semiconductor is feasible from a conventional
ferromagnet when the interface resistance between the semiconductor
and the ferromagnet is sufficiently large, as recently demonstrated by
using Fe Schottky contacts in Ref.~\onlinecite{lou:nphys:v3:2007}.
Spin injection detected via the GMR effect in nanotubes contacted to
ferromagnets have also been reported.\cite{Alphenaar:jap89}

Graphene is clearly an interesting candidate for spintronics
applications since the carrier concentration can be controlled by gate
voltages. Also, it has a very weak spin-orbit interaction, leading to
the possibility of relatively long spin flip
lengths.\cite{Huertas-Hernando:prb74,Huertas-Hernando:proc-graphene-06}
In a recent experiment on spin injection in single layer graphene the
spin flip length is found to be $l_\text{sf} \approx
\unit{1}{\micro\metre}$ at room temperature in dirty
samples.\cite{vanWees:2007} Cleaner samples are expected to have even
longer spin flip lengths.

We explore another possibility of spin dependent transport by
envisioning that graphene is put in close proximity to a magnetic
insulator.  Via the magnetic proximity effect, exchange splittings
will be induced in graphene. Strong proximity induced exchange
splittings due to ferromagnetic insulators have been observed at
$\text{EuO}/\text{Al}$ interfaces.~\cite{Tedrow:prl:v56:p1746,
  Tkaczyk:prl:v61:p1253, Hao:prl:v67:p1342} The effect was attributed
to the non-vanishing overlap between the wave functions of the
localized moments in the magnetic insulator and the itinerant
electrons in the metal.\cite{Roesler:SPIE:1994} 
The electronic wave functions can be described by atomic-like wave
functions at the surface of thin $\text{Al}$
films.\cite{painter:prb:v17:p662} %
The spatial range is similar for the atomic wave functions in Al and
graphene, so we expect the overlap between the localized moments and
itinerant electrons in graphene at $\text{EuO}/\text{graphene}$
interfaces to induce exchange interactions comparable to those
observed for $\text{EuO}/\text{Al}$. Based on the results reported in
Refs.~\onlinecite{Roesler:SPIE:1994,
  Tokuyasu:prb:v38:p8823,Hao:prl:v67:p1342, Tkaczyk:prl:v61:p1253} we
roughly estimate that exchange splittings in graphene due to the
ferromagnetic insulator EuO could be of the order of $\unit{5}{\meV}$
(see Appendix~\ref{sec:app-euo} for details).

In this paper, we show that proximity induced splittings can be
observed in the tunneling conductance associated with a tunable
barrier created by the ferromagnetic insulator gate on top of
graphene. First, for highly doped barriers, we demonstrate that
the splitting $\Delta$ can be directly observed from the transmission
resonances in the conductance.\cite{Katsnelson2006nphys,
  cheianov-2007-315,pereirajr-2007} Moreover, for low doping of the
barrier we show that the spin polarization of the tunneling current,
directly related to the spin splitting $\Delta$, increases with
increasing length of the barrier. The spin polarization can be studied
by magnetoresistance (MR) measurements in a spin valve device where
two magnetic gates are deposited in series. Such MR measurements could
also allow to independently determine the induced spin splitting
$\Delta$.

This paper is organized as follows: We present a model of a magnetic
gate in Sec.~\ref{sec:model}. Section~\ref{sec:tunprob} reminds the
reader of the results obtained in
Refs.~\onlinecite{Katsnelson2006nphys}
and~\onlinecite{tworzydlo:prl:v96:p246802} for the conductance of a
square barrier in graphene. Then we discuss how to obtain analytical
expressions for the conductance both far from and close to the Dirac
point. We extend the spinless situation to a spin dependent barrier
with an exchange splitting $\Delta$ between the two spin channels in
Sec.~\ref{sec:spidepbar}. First, we discuss the possibilities for
extracting the splitting $\Delta$ directly from the conductance of a
single highly doped barrier. Second, we study the dependence of the
current spin polarization on the barrier height and
length. Section~\ref{sec:mag} discusses the MR effect in a double
barrier spin valve device and discusses how it can be used to extract
$\Delta$. Finally, our conclusions are in Sec.~\ref{sec:concl}.

\section{Model}
\label{sec:model}

\begin{figure}[htbp]
  \centering
  \includegraphics{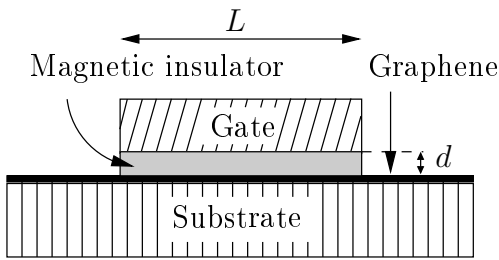}
  \caption{A ferromagnetic insulator on top of graphene induces an
    exchange splitting in graphene. A metallic gate on top of the
    insulator controls the electrostatic potential.}
  \label{fig:modeldescription}
\end{figure}

A possible way to construct a ferromagnetic gate is to deposit a
magnetic insulator, such as EuO, on top of a graphene sample with a
metallic gate above it (see Fig.~\ref{fig:modeldescription}). So far,
experimental efforts have focused on depositing non-magnetic gates on
graphene.\cite{ozyilmaz:2007,huard:prl98:y2007} The presence of a
magnetic insulator will induce an exchange splitting in graphene. The
normal metal gate allows to control the Fermi level locally,
\emph{i.e.} to create a tunable barrier in graphene. In this way, both
control of the charge and spin carrier concentrations can be achieved.

We assume in this paper that the normal metal gate induces a sharp
potential barrier below it. This is a reasonable assumption provided
the distance $d$ between the gate and the graphene layer is much
shorter than the Fermi wavelength $\lambda_F$, which is relatively
long in graphene, $\lambda_F \approx
\unit{50-100}{\nm}$.\cite{Katsnelson2006nphys} Recently, a method for
manufacturing top gates where the distance from the gate to the
graphene layer is of the order of $\lambda_F$ has been
demonstrated.\cite{ozyilmaz:2007} Observation of resonance
effects due to sharp potential steps therefore seems feasible in
graphene.

The exchange interaction between the localized magnetic moments in the
ferromagnetic insulator and the spins of the electrons creates an
additional spin dependent offset of the barrier potential, leading to
the possibility of spin dependent tunneling. We estimate in
Appendix~\ref{sec:app-euo} that the exchange splitting due to the
magnetic insulator EuO can be around $\unit{5}{\meV}$. Here we assume
that the exchange interaction is not affected by the gate voltage of
the top metallic gate.

\section{Tunneling probability}
\label{sec:tunprob}

\begin{figure}[htbp]
  \includegraphics{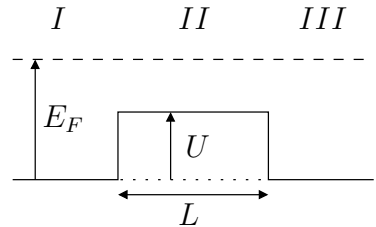}
  \caption{Square barrier of length L}
  \label{fig:square-barrier}
\end{figure}

For completeness, we first review the results for tunneling through a
square barrier in graphene, and follow the derivation in
Refs.~\onlinecite{Katsnelson2006nphys}
and~\onlinecite{tworzydlo:prl:v96:p246802}.  We will later extend this
discussion to a spin dependent barrier. The charge carriers we
consider are Dirac quasi-particles, described by the
Hamiltonian~\eqref{eq:graphene-low-energy-hamiltonian}. These
quasi-particles originate from reservoirs to the left and to the right
of the ballistic graphene sample.  $E_F$ is the Fermi energy measured
with respect to the Dirac point of the undoped graphene layer. At zero
temperature, the transport properties are governed by quasi-particles
that approach a square barrier of height $U$ and length $L$ (see
Fig.~\ref{fig:square-barrier}) at energy $E_F$.  We assume ballistic
transport across the barrier, and also that the spin flip length
$l_\text{sf}$ is much longer than the other length scales of the
problem. Given the values for $l_\text{mfp}$ and $l_\text{sf}$
reported for graphene,\cite{vanWees:2007,
  Huertas-Hernando:proc-graphene-06, Novoselov2004sci306} this regime
should be realistic.

The Hamiltonian~\eqref{eq:graphene-low-energy-hamiltonian} has the
following plane wave solutions in regions $I$, $II$, and $III$ of
Fig.~\ref{fig:square-barrier},
respectively:\cite{Katsnelson2006nphys,tworzydlo:prl:v96:p246802}
\begin{align}  
  \label{eq:wavefunction-region-III}
  \psi_{(I)} &= 
  \left[ 
    \begin{pmatrix}
      1 \\ 
      \alpha e^{i \theta}%
    \end{pmatrix}
    e^{i k_x x} + r 
    \begin{pmatrix}
      1 \\ 
      - \alpha e^{-i \theta}%
    \end{pmatrix}
    e^{-i k_x x} \right] e^{i k_y y}, \\
  \psi_{(II)} &=
  \left[ a 
    \begin{pmatrix}
      1 \\ 
      \beta e^{i \phi}%
    \end{pmatrix}
    e^{i q_x x} + b 
    \begin{pmatrix}
      1 \\ 
      -\beta e^{-i \phi}%
    \end{pmatrix}
    e^{-i q_x x} \right] e^{i q_y y}, \\
  \psi_{(III)} &=
  \begin{pmatrix}
    1 \\
    \alpha e^{i \theta}%
  \end{pmatrix}
  e^{i k_x \left( x - L \right)} e^{i k_y y}.
\end{align}
The momentum of the incident particle makes an angle $\theta =
\arctan{k_y/k_x}$ with the $x$ axis. The angle of refraction, i.e.
the corresponding angle inside the barrier, is $\phi =
\arctan{q_y/q_x}$. We consider only elastic scattering at the
interfaces and define
\begin{align}
  k_F &
  \equiv \left( k_x^2 + k_y^2 \right)^{1/2} 
  = (\hbar v)^{-1} |E_F| \\
  \text{and} \quad 
  q_F &
  \equiv \left( q_x^2 + q_y^2 \right)^{1/2} 
  = (\hbar v)^{-1} |E_F-U|.
\end{align}
The parameters $\alpha = \signum{E_F}$ and $\beta = \signum{E_F-U}$
determine the wave function in the corresponding regions as either
electron like (positive sign) or hole like (negative sign).
Translational invariance in the transverse ($y$) direction implies
conservation of transverse momentum:
\begin{equation}
  \label{eq:transv-cont}
  k_y = q_y 
  \quad \Rightarrow \quad
  k_F \sin{\theta} = q_F \sin{\phi}.
\end{equation}
It is convenient to introduce the dimensionless variable
\begin{align}
  \label{eq:xi-def}
  \xi &= \frac{E_F - U}{E_F} 
\end{align}
as a measure of the gate voltage $U$. $\xi=1$ corresponds to the case
of no barrier. Throughout the rest of the paper we will make the
substitution $u = \sin{\theta}$, and we recall that by definition
$\alpha k_F = E_F / \hbar v$ and $\beta q_F = (E_F-U)/ \hbar v$.

Matching the wave functions at the interfaces, $\psi_{(I)}(x=0) =
\psi_{(II)}(x=0)$ and $\psi_{(II)}(x=L) = \psi_{(III)}(x=L)$, and
solving for $t$ gives the transmission probability $T \equiv |t|^2$
for a given incoming angle
$\theta$:\cite{Katsnelson2006nphys,tworzydlo:prl:v96:p246802}
\begin{equation}
  \label{eq:transm-prob-angle}
  \begin{split}
    T(u)
    &= 
    \frac{ 
      \left(\xi^2 - u^2 \right) \left( 1 - u^2 \right)
    }{
      \left(\xi^2 - u^2 \right) \left( 1 - u^2 \right)
      + u^2 \left(1 - \xi \right)^2 \sin^2{(q_x L)}
    },
  \end{split}
\end{equation}
where
\begin{equation}
  \label{eq:qxL}
  q_x L
  = k_F L \sqrt{ \xi^2 - \sin^2{\theta} }.
\end{equation}
Both $t$ and $T$ are invariant under the transformation $k_y \to -k_y$
as a consequence of the continuity condition~\eqref{eq:transv-cont}.

In a real device, the sample has a finite width $W$. The allowed
incoming angles $\theta$ are therefore determined by the channel index
$n$, due to the quantization of the transverse modes.  This
quantization condition is, for the infinite mass boundary condition,
$k_y \to k_n = \left( n + \frac{1}{2} \right)/W$, where $n$ are
integers in the range $0$ to $N_\text{max} = \lfloor k_F
W/\pi-1/2\rfloor$, and the transverse states are superpositions of
states with positive and negative
$k_y$.\cite{berry:prsla:v412:p53,tworzydlo:prl:v96:p246802} Provided
that the transverse momentum is conserved across the barrier
interfaces, Eq.~\eqref{eq:transm-prob-angle} is valid for systems of
both finite and infinite width.\cite{tworzydlo:prl:v96:p246802}

\begin{figure}[htbp]
  \centering
  \includegraphics[width=\columnwidth]{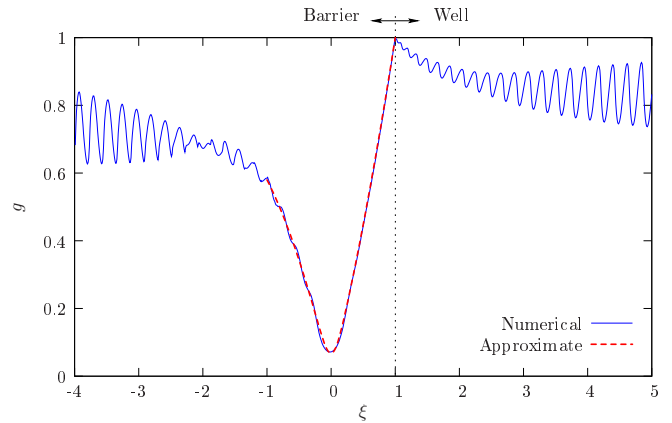}
  \caption{Conductance $g = G/\mathcal{G}_0$ as a function of $\xi =
    (E_F-U)/E_F$ normalized to one spin channel when $k_F L = 14$.
    The solid (blue) line shows the numerical result
    using~\protect\eqref{eq:GlargeW} while the dashed (red) line is
    computed using the
    approximation~\protect\eqref{eq:cond-dirac-vicinity}. $\mathcal{G}_0
    = 2e^2 k_F W /(h \pi)$.}
  \label{fig:transmission-integral-barrier}
\end{figure}

The conductance through the barrier for each spin independent channel
is given in the Landauer-B\"{u}ttiker formalism as 
\begin{equation}
  G = g_v \frac{e^2}{h} \sum_{n = 0}^{N_\text{max}} T_n,
\end{equation}
where $g_v = 2$ is the valley degeneracy and $T_n$ is the transmission
probability, Eq.~\eqref{eq:transm-prob-angle}, for a given transverse
channel $k_n$.  When the number of channels $N$ becomes large, i.e.
$k_F W \gg 1$, we can replace the summation over channels with an
integration over transverse momenta, such that the conductance becomes
\begin{equation}
  \label{eq:GlargeW}
    G = \mathcal{G}_0 \int_0^{1} \form{u} T(u)
    = {\mathcal{G}_0} g
\end{equation}
with $\mathcal{G}_0$ defined as
\begin{equation}
  \label{eq:conducatance-normalization}
  \mathcal{G}_0  
  = \frac{2e^2}{h} \frac{ k_F W}{\pi}
  .
\end{equation}
The dimensionless conductance $g$ in~\eqref{eq:GlargeW} is plotted in
Fig.~\ref{fig:transmission-integral-barrier} as a function of the
dimensionless gate voltage $\xi$.

From Eq.~\eqref{eq:qxL}, we see that the longitudinal momenta in the
barrier region, $q_x$, can be either purely real ($\xi^2 > u^2$) or
purely imaginary ($\xi^2 < u^2$), corresponding to propagating and
evanescent modes, respectively.\cite{tworzydlo:prl:v96:p246802} The
contribution to the conductance from the evanescent modes becomes
dominant around $\xi = 0$, and the scaling of the conductance with
length at this point resembles that of a diffusive
system.\cite{tworzydlo:prl:v96:p246802,prada:prb:v75:p113407:2007} For
$|\xi| < 1$, the conductance~\eqref{eq:GlargeW} can be split into the
contributions from propagating and evanescent modes:
\begin{equation}
  \label{eq:cond_contribs}
  \begin{split}
    g 
    &= \int_0^{|\xi|} \form{u} T(u) + \int_{|\xi|}^{1} \form{u} T(u) \\
    &= g_\text{prop} + g_\text{evan},
  \end{split}
\end{equation}
from which it is readily seen that the evanescent modes dominate in
the region near $\xi = 0$ as long as $T(u) > 0$ for at least some $u >
|\xi|$, (see Appendix~\ref{sec:app-limits} for details). For $k_F L \gg
1$ and setting $\xi = 0$ in Eq.~\eqref{eq:transm-prob-angle}:
\begin{equation}
  \label{eq:Tevanesc}
  \begin{split}
    T(u) 
    &
    \approx 
    \frac{
      1
    }{
      \cosh^2{(k_F L u)}
    }.
  \end{split}
\end{equation}
This corresponds to the limit $N_\text{max} \gg W/L$ in
Ref.~\onlinecite{tworzydlo:prl:v96:p246802}. Upon insertion
of~\eqref{eq:Tevanesc} into the integral~\eqref{eq:GlargeW}, we find
that the conductance at the Dirac point is inversely proportional to
the system length:
\begin{equation}
  \label{eq:GUequalsE}
  g \approx \frac{1}{k_F L}.
\end{equation}
This corresponds to the so-called minimal conductivity $g_s G \times
L/W = g_s g_v e^2/ (h \pi)$,\cite{tworzydlo:prl:v96:p246802} $g_s=2$
being the spin degeneracy. 

For $|\xi| < 1$ and $k_F L \gg 1$ we can approximate the conductance
by the expression
\begin{equation}
  \label{eq:cond-dirac-vicinity}
  g \approx
  (a_1 + a_2 \xi) |\xi|
  +
  \frac{1}{k_F L}
  \exp{(- k_F L |\xi|)}
  ,
\end{equation}
with $a_1 = 0.79$ and $a_2 = 0.21$ (see
Appendix~\ref{sec:length-depend-struct} for details and
Fig.~\ref{fig:transmission-integral-barrier} for a comparison with the
exact solution). Equation~\eqref{eq:cond-dirac-vicinity} reduces
to~\eqref{eq:GUequalsE} when $\xi \to 0$.

For $|\xi| > 1$, corresponding to a well or a large barrier, 
only propagating modes contribute, and we would expect to see
resonances in the conductance due to quasi-bound~\cite{Silvestrov2007}
states in the barrier region.  In the limit $|\xi| \gg 1$, using that
$u^2 \leq 1$, the tunneling probability~\eqref{eq:transm-prob-angle}
becomes
\begin{equation}
  \begin{split}
    T(u) \approx 
    \frac{
      1-u^2
    }{
      1-u^2 \cos^2{(k_F L \xi)} 
    },
  \end{split}
\end{equation}
resulting in the expression 
\begin{equation}
  \label{eq:GlargeU}
  \begin{split}
    g &\approx
    \frac{
      |\cos(k_F L \xi)| 
      - \sin^2(k_F L \xi) \arctanh(|\cos(k_F L \xi)|)
    }{
      |\cos(k_F L \xi)|^3
    }
  \end{split}
\end{equation}
for the dimensionless conductance (see Appendix~\ref{sec:app-limits}
for details). The period of $g$ as a function of $\xi$ is $\pi/k_F
L$. Also $g$ oscillates between $2/3$ and $1$. The transmission
probability analogous to~\eqref{eq:transm-prob-angle} for a square
barrier in a non-chiral two dimensional
system,\cite{Katsnelson2006nphys}
\begin{equation}
  \label{eq:transm-prob-2deg}
  \begin{split}
    T_\text{non-chiral} 
    & = \frac{
      4 (\xi^2-u^2) (1-u^2) 
    }{
      4 (\xi^2-u^2) (1-u^2)
      + (1 - \xi^2)^2 \sin^2{(q_x L)}
    },
  \end{split}
\end{equation}
also gives oscillation with the same periodicity, but in this case the
conductance oscillates between $0$ and $1$. The fact that the
conductance given by Eq.~\eqref{eq:GlargeU} oscillates between $2/3$
and $1$ is due to the perfect tunneling of carriers near normal
incidence in graphene. Another difference between graphene and a
non-chiral system is that the transmission probability of the latter,
\eqref{eq:transm-prob-2deg} is symmetric around $\xi = 0$, while the
transmission probability for graphene, \eqref{eq:transm-prob-angle},
depends also on the sign of $\xi$ through the $(1-\xi)$-factor in the
denominator. The asymmetry for the case of graphene can readily be
seen in Fig.~\ref{fig:transmission-integral-barrier}.

\section{Spin dependent barrier}
\label{sec:spidepbar}

We now turn to a situation where the two spin channels see barriers of
different heights, i.e. the bottom of the conduction band at the
barrier is shifted differently according to spin. Such a shift can
arise through a Zeeman interaction due to an in-plane magnetic field
or exchange field. 

\begin{figure}[htbp]
  \includegraphics{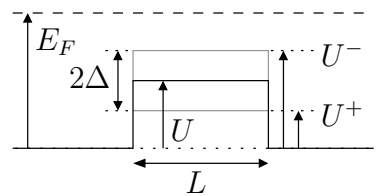}
  \caption{Ferromagnetic proximity effect splits the barrier according
    to spin such that $U^\pm = U \mp \Delta$.}
  \label{fig:barrier-split}
\end{figure}

The exchange term $\Delta$ splits the system into two separate
subsystems according to spin. For an external magnetic field $B$, the
splitting is given by $\Delta \approx 2 \mu_B B$. We introduce the
spin dependent variables
\begin{equation}
  \label{eq:xi_pm} 
  \xi^\pm = \xi \pm \delta = \frac{E_F - U}{E_F} \pm \frac{\Delta}{E_F},
\end{equation}
where $\pm$ denotes spins parallel ($+$) or anti parallel ($-$) to the
exchange field (see Fig.~\ref{fig:barrier-split}). In the following we
will let $g^{+(-)}$ denote the spin resolved conductance for spins
parallel (anti-parallel) to the exchange field. Assuming no spin flip,
$l_\text{sf} \gg L$, the total conductance $g_T$ across the barrier is
given by the sum:
\begin{equation}
  \label{eq:total-conductance}
  g_T = g^+ + g^- = g(\xi^+) + g(\xi^-).
\end{equation}

Because $\Delta/B \approx \unit{5.8 \times 10^{-2}}{\meV\per\tesla}$,
a direct interaction of the spins with an external magnetic field
gives only a very weak effect (about $\unit{1}{\meV}$ per
$\unit{20}{\tesla}$), and one will have to rely on more indirect
effects to observe such spin splittings.

We propose depositing a ferromagnetic insulator, \emph{e.g.}
$\text{EuO}$, on top of the graphene sample to induce an exchange
splitting in graphene. A normal gate on top of the insulator allows to
control the Fermi level in the same region. The resulting potential
profile is sketched in Fig.~\ref{fig:barrier-split}. A rough estimate
suggests that the splitting energy can be of order $\Delta \approx
\unit{5}{\meV}$ at EuO/graphene interfaces (see
Appendix~\ref{sec:app-euo}).

\begin{figure}[htbp]
  \centering
  \includegraphics[width=\columnwidth]{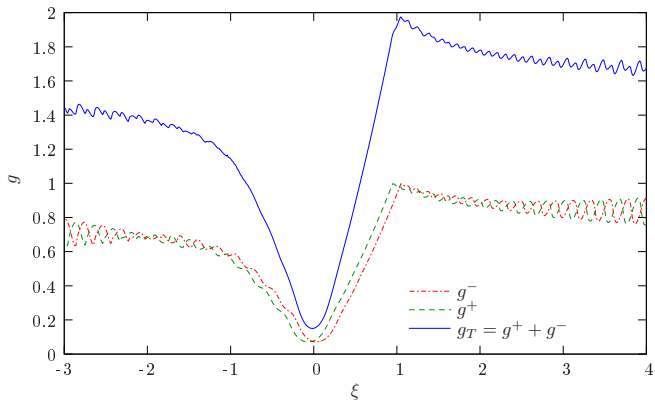}
  \caption{Spin resolved conductance through a square barrier for $k_F
    L = 14$ and $\delta = \Delta/E_F = 0.05$. The normalization of
    conductance is chosen as in
    Fig.~\protect\ref{fig:transmission-integral-barrier} to correspond
    to $g(1) = 1$ for each spin channel.}
  \label{fig:splittedbarrier}
\end{figure}

As can be seen from Fig.~\ref{fig:splittedbarrier} the effect of the
splitting is simply to shift the conductance of each spin channel with
respect to the other, leading to a broadening of the dip in the total
conductance $g_T$ near the Dirac point $\xi=0$. To be able to observe
the splitting directly in the $g_T$ near the Dirac point, the
magnitude of the splitting must be larger than the width of the dip of
each spin resolved conductance, $g^{+(-)}$. A measure $w = (k_F
L)^{-1}$ of the width of the dip is discussed in
Appendix~\ref{sec:length-depend-struct}, leading to the condition
\begin{equation}
  \label{eq:observation-condition}
  L > \frac{\hbar v}{\abs{\Delta}}.
\end{equation}
for observation of the splitting directly in $g_T$ at the Dirac point.
However, the broadening of the dip due to a spin splitting would be
difficult to distinguish from a broadening due to other effects.

\begin{figure}[htbp]
  \centering
  \includegraphics[width=\columnwidth]{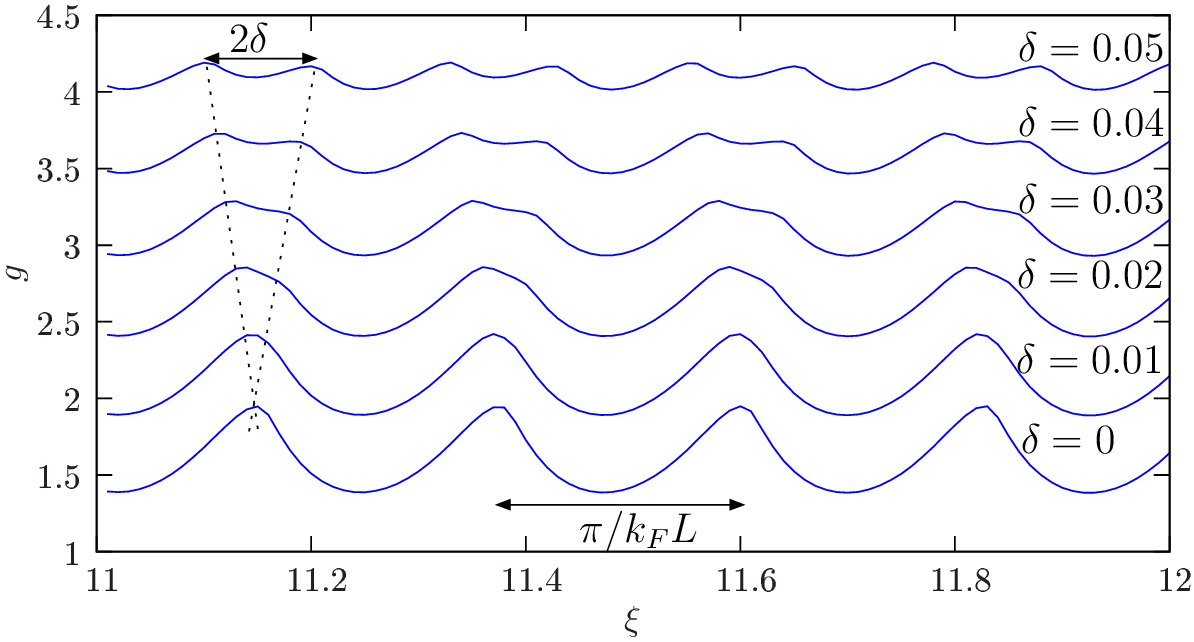}
  \caption{Total conductance $g_T = g^+ + g^-$ when $\abs{\xi} \gg 1$
    for a range of different splittings $\delta = \Delta/E_F$. For
    clarity the curves are shifted upwards in steps of $0.5$
    with increasing $\delta$.}
  \label{fig:splitsuccession}
\end{figure}

From Fig.~\ref{fig:splittedbarrier} it is apparent that the spin
splitting has a more dramatic effect on the total conductance $g_T$ at
large barrier doping, since due to the transmission resonances,
$g^{+}$ and $g^{-}$ can differ substantially at a given $\xi$. The
asymptotic expression~\eqref{eq:GlargeU} for $|\xi| \gg 1$ implies
that $g_T$ has periodicity $\pi/k_F L$ in $\xi$ for $\delta=0$, as
shown at the bottom of Fig.~\ref{fig:splitsuccession}. With increasing
$\delta$, each peak of $g_T$ gradually splits into two spin resolved
peaks. The splitting measured from the conductance $2 \delta$ equals $
2 \Delta/E_F$ (see Fig.~\ref{fig:splitsuccession}), so in principle
$\Delta$ can be determined directly from the total conductance across
the barrier in this way.

On the other hand, it is also possible to study the splitting by
examining the spin polarization across the barrier.  

\begin{figure}[htbp]
  \centering
  \includegraphics[width=\columnwidth]{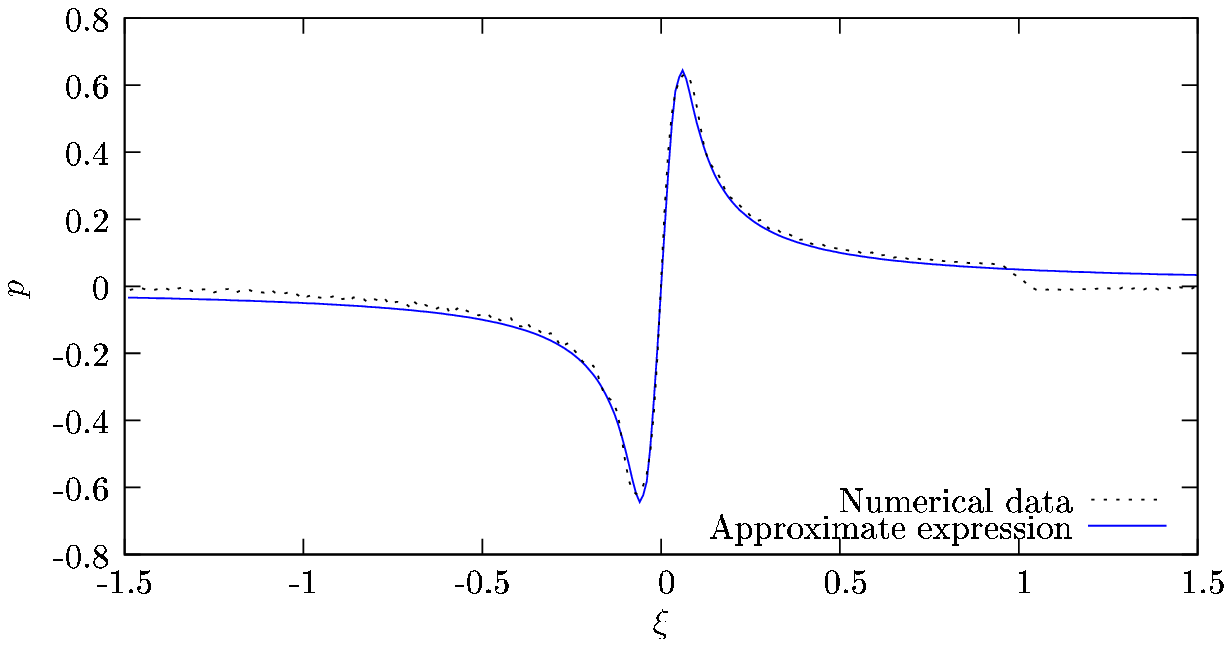}
  \caption{The polarization $p$ from the
    approximation~\protect\eqref{eq:cond-dirac-vicinity} compared to
    the exact numerical result obtained directly from
    Eq.~\eqref{eq:GlargeW}.  Both plots are for $k_F L = 14$ and
    $\delta = 0.05$.}
  \label{fig:polarisationapprox}
\end{figure}

We define a normalized spin polarization $p$ along the
direction of the exchange field as
\begin{equation} 
  \label{eq:polarisation-def}
  p = \frac{g^{+} - g^{-}}{g^{+} + g^{-}}.
\end{equation}
Inserting the approximate expression for the conductance from
Eq.~\eqref{eq:cond-dirac-vicinity} and comparing to exact numerical
calculations, we find good agreement in the whole region $|\xi| < 1$
(see Fig.~\ref{fig:polarisationapprox}).

\begin{figure}[htbp]
  \centering
  \includegraphics[width=\columnwidth]{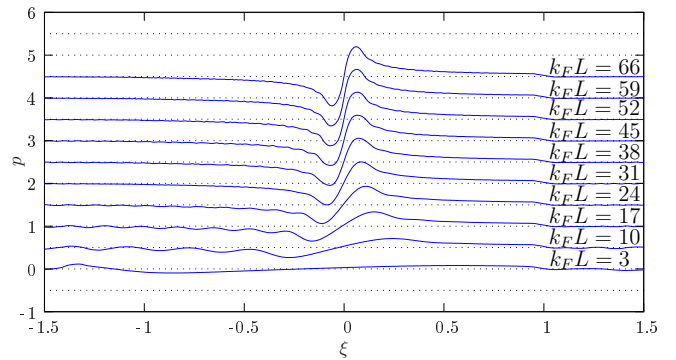}
  \caption{Polarization $p$ as a function of $\xi = (E_F-U)/E_F$ for
    different barrier lengths $L$ with $\delta = 0.05$.}
  \label{fig:polarisation}
\end{figure}

Equation~\eqref{eq:cond-dirac-vicinity} implies that the polarization becomes
more pronounced with increasing barrier length $L$ (see
Fig.~\ref{fig:polarisation}), owing to the fact that the evanescent
modes are increasingly suppressed as $L$ increases.

\section{Magnetoresistance}
\label{sec:mag}

\begin{figure}[htbp]
  \includegraphics{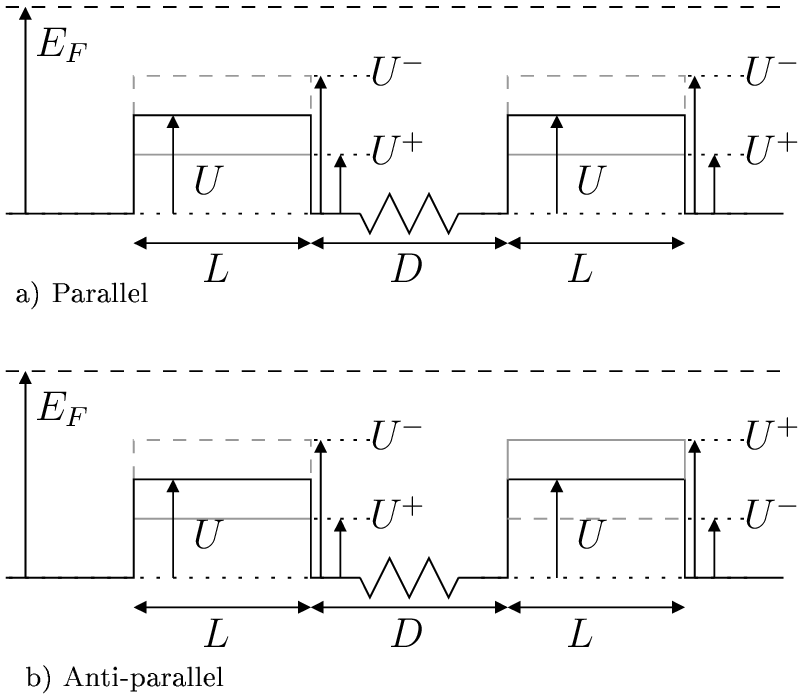}
  \caption{Measuring tunneling magnetoresistance by placing two short
    barriers a distance $D \gg l_\text{mfp}$ apart.}
  \label{fig:magnetoresistance}
\end{figure}

Placing two magnetic gates a distance $D$ apart in the graphene sample
is a possible way to probe the polarization $p$ in
Eq.~\eqref{eq:polarisation-def}. We assume either that $D$ is much
larger than the mean free path $l_\text{mfp}$ (but still much shorter
than $l_\text{sf}$), or that the experimental setup is realized as a
three-terminal experiment, where the middle terminal completely
randomizes the momenta between the two barriers (see
Fig.~\ref{fig:magnetoresistance}).

Assuming that no spin flip processes take place in the sample, the
conductance for each spin channel is found by treating the two
barriers as resistors connected in series. Arranging the
magnetizations of the ferromagnetic barriers parallel or anti-parallel
to each other, gives different conductances $g_{\Uparrow \Uparrow}$
and $g_{\Uparrow \Downarrow}$, corresponding to the two situations in
Fig.~\ref{fig:magnetoresistance}, respectively. We study the
polarization using the ``pessimistic'' definition of the
magnetoresistance: $MR = \left( g_{\Uparrow \Uparrow}-g_{\Uparrow
    \Downarrow} \right) / g_{\Uparrow
  \Uparrow}$. 
For the general case of different left $(L)$ and right $(R)$ barriers,
we obtain
\begin{equation} 
  \label{eq:mr}
  \begin{split}
  MR= 
  \frac{
    4 p_L g_L p_R g_R 
  }{
    \left( g_L + g_R \right)^2
    - \left( p_L g_L - p_R g_R \right)^2
  },
  \end{split}
\end{equation}
assuming that the resistance of the region $D$ between the barriers is
negligible compared to the typical resistances of the barriers. For
clarity we have suppressed the subscript $T$ denoting total
conductance of the left (right) barrier: $g_{L(R)} \equiv g_{L(R)}^{+}
+ g_{L(R)}^{-}$.

\begin{figure}[htbp]
  \centering
  \includegraphics[width=\columnwidth]{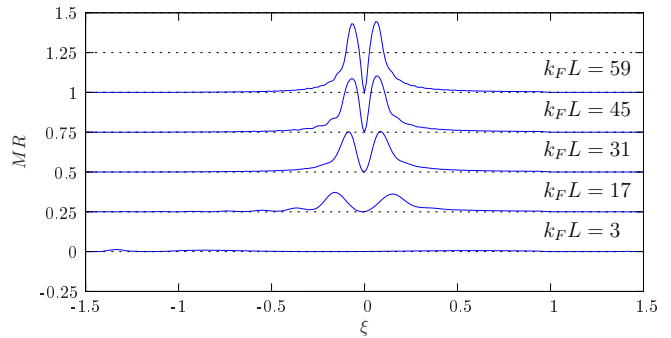}  
  \caption{Magnetoresistance for two barriers of equal height. The
    curves are shifted upward in steps of $0.25$ for clarity.}
  \label{fig:mrsym}
\end{figure}

\begin{figure}[htbp]
  \centering
  \includegraphics[width=\columnwidth]{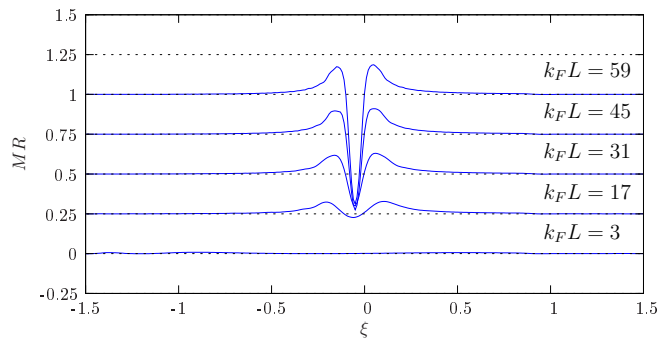}
  \caption{Same as in Fig.~\protect\ref{fig:mrsym}, with one barrier
    being lower than the other ($|\xi_L - \xi_R| = 0.1$).}
  \label{fig:mrasym}
\end{figure}

For identical barriers, $MR$ reduces to the simple expression:
\begin{equation}
  \label{eq:mr-symmetric}
  MR = p^2.
\end{equation}
The combination of
Eqs.~\eqref{eq:cond-dirac-vicinity},~\eqref{eq:polarisation-def}
and~\eqref{eq:mr} allows to experimentally determine $\Delta$ from
magnetoresistance measurements.  The change of sign in the
polarization, shown in Fig.~\ref{fig:polarisation}, is directly
related to the relative shift of the conductances corresponding to
each spin channel. The coefficient $MR$ is proportional to $p^2$,
which produces the double peak structure seen in Figs.~\ref{fig:mrsym}
and~\ref{fig:mrasym}. The condition for observing MR effects is also
given by Eq.~\ref{eq:observation-condition}, $L > \hbar
v/\abs{\Delta}$. However, since the MR signal is only sensitive to the
spin degree of freedom, we expect MR experiments to be a more direct
probe of a spin induced splitting. Any broadening of the dip
introduced by sources other than $\Delta$ will also be less important,
since the polarization $p$ changes sign around $\xi=0$.

For a splitting of $\Delta = \unit{5}{\meV}$, the condition in
Eq.~\eqref{eq:observation-condition} gives $L > \unit{110}{\nm}$ (or
equivalently $k_F L > 20$). As can be seen in
Fig.~\ref{fig:polarisation}, the features in the polarization becomes
sharper when increasing the length above this value. This also
translates into a clearer signal in the magnetoresistance, which is
plotted in Fig.~\ref{fig:mrsym} and~\ref{fig:mrasym} for barriers of
equal and unequal heights, respectively.

Finally, even if the top gate creates a smooth tunable barrier, far
from the the perfectly square potential discussed here,
magnetoresistance measurements should still provide an experimental
demonstration of proximity induced ferromagnetism in graphene, as the
magnetic insulator still creates a sharp splitting of the spin up and
spin down states in the region underneath the magnetic insulator.  The
exact dependence of the polarization $p$ on the splitting $\Delta$
may be different in this case than the one presented here.

\section{Conclusions}
\label{sec:concl}

We suggest using magnetic insulators deposited on top of graphene to
create ferromagnetic graphene.  The exchange interaction between
electrons in graphene and the localized magnetic moments in the
insulator will give rise to a proximity induced exchange splitting
$\Delta$. We have estimated that the graphene exchange splitting due
to the magnetic insulator EuO in close proximity can be around $\Delta
= \unit{5}{\meV}$.

We have studied how the conductance of a square barrier in graphene is
modified by the presence of a ferromagnetic insulator. We show that
for large barriers or deep wells, $\abs{\xi} \gg 1$, the splitting
$\Delta$ can be determined directly from the total conductance across
the barrier, provided that the barrier is sharp enough for
transmission resonances to appear. For a barrier of length $L > \hbar
v/\abs{\Delta}$, where $v$ is the Fermi velocity of the charge
carriers in graphene, $\Delta$ should be observable in the
polarization of the tunneling current near the Dirac point of the
barrier, irrespectively of whether the barrier is smooth or sharp.

Demonstration of proximity induced ferromagnetism in graphene should
be possible through magnetoresistance measurements both for smooth
and sharp barriers.
\vspace{0.5cm}
 
Note added: After completion of this work we became aware of a related
work by Y.G. Semenov et al.,\cite{Semenov:2007} where a similar system
with a magnetic gate is considered.  Their work discusses the
possibility of a spin FET which feasibility relies on variations of
the spin splitting across the sample of the same order of magnitude as
our estimate for the splitting itself.

\begin{acknowledgments}
  We thank G.~E.~W.~Bauer, M.~F.~Craciun, F.~Guinea, Yu.~V.~Nazarov,
  E.~Prada, and S.~Russo for discussions. This work was supported by
  EC Contract IST-033749 ``DynaMax'' and The Research Council of
  Norway through grant no. 167498/V30 and no. 162742/V30.
\end{acknowledgments}

\appendix

\section{Estimation of exchange splitting at $\text{EuO}/$graphene
  interfaces}
\label{sec:app-euo}

Experiments on depairing at $\text{EuO}/\text{Al}$ interfaces suggest
that the superconducting quasi-particles of $\text{Al}$ experience an
exchange field due to the $\text{Eu}^{2+}$
moments.\cite{Tedrow:prl:v56:p1746,Tokuyasu:prb:v38:p8823} This
interaction is short ranged; essentially only the nearest layer of
$\text{Eu}^{2+}$ ions contributes. It has be shown that the exchange
interaction between $\text{Eu}^{2+}$ ions and charge carriers can be
described as a Zeeman
splitting~\cite{Tedrow:prl:v56:p1746,Roesler:SPIE:1994,
  Tokuyasu:prb:v38:p8823,Hao:prl:v67:p1342, Tkaczyk:prl:v61:p1253}
\begin{equation}
  \label{eq:appendix-estimation-1}
  \Delta \approx c J \langle S_z \rangle,
\end{equation}
where $c$ is the fractional density of $\text{Eu}^{2+}$ ions to that
of itinerant electrons in Al at the interface, $J$ is the spatial
average of the exchange integral and $\langle S_z \rangle$ is the
average spin of $\text{Eu}^{2+}$ ions at a given temperature.

Perpendicular to the surface of thin $\text{Al}$ films, the electronic
wave functions can be well approximated by atomic-like wave
functions.~\cite{painter:prb:v17:p662}. The spatial range of an atomic
wave function is determined by the ratio
$Z/n$,\cite{brehm:mullin:1989} where $Z$ is the atomic number and $n$
is the energy level.  Since this ratio is approximately the same for
the $3s$ and $3p$ orbitals in $\text{Al}$ ($Z/n = 13/3 \approx 4.3$)
and the $2p$ orbitals in graphene ($Z/n = 6/2 = 3$), we expect the
overlap between the wave functions of localized moments and itinerant
electrons at $\text{EuO}/\text{graphene}$ interfaces to be comparable
to those for $\text{EuO}/\text{Al}$. Accordingly, we assume that the
exchange interaction between $\text{Eu}^{2+}$ ions and itinerant
electrons to be the same at $\text{EuO}/\text{Al}$ and
$\text{EuO}/\text{graphene}$
interfaces. Ref.~\onlinecite{Roesler:SPIE:1994} reports the value $J =
\unit{15}{\meV}$ for $\text{Eu}/\text{Al}$ interfaces, which also
agrees with the exchange energy $h_\text{ex} = \unit{0.1}{\meV}$
estimated in Ref.~\onlinecite{Tokuyasu:prb:v38:p8823}.

Using a nearest neighbor distance in graphene of
$\unit{1.42}{\angstrom}$,\cite{wallace:pr:v71:p622:y1947} we obtain
for the areal density of itinerant electrons $n_\text{C} \approx
\unit{40}{\nm^{-1}}$. Similarly, the areal density of
$\text{Eu}^{2+}$-ions at the surface of $\text{EuO}$ is
$n_{\text{Eu}^{2+}} \approx \unit{4}{\nm^{-1}}$. Together this gives
$c = n_{\text{Eu}^{2+}} / n_\text{C} \approx 10^{-1}$.

The temperature dependence of the average spin of $\text{Eu}^{2+}$
ions in $\text{EuO}$ is calculated in
Ref.~\onlinecite{Nolting:prb:v35:p7025}, showing that $3.5 \geq
\langle S_z \rangle \geq 3$ for $0 < T < \unit{30}{\kelvin}$.

Collecting all together we arrive at the estimate
\begin{equation}
  \Delta \approx \unit{5}{\meV}
\end{equation}
for the exchange splitting in graphene due to $\text{EuO}$. We stress
that this is a very rough estimate which needs to be tested
experimentally.

\section{Limiting cases for the conductance}
\label{sec:app-limits}

\subsubsection{Large potential}

For large barriers or deep wells, $|E_F-U| \gg |E_F|$, $\xi^{-1} \ll
1$.  The transmission probability~\eqref{eq:transm-prob-angle} then
becomes
\begin{equation}
  \label{eq:limits1}
  \begin{split}
    T(u) 
    &
    \approx
    \frac{
      1- u^2
    }{
      1 - u^2 \cos^2{\left(k_F L \xi \right)} 
    }.  
  \end{split}
\end{equation}
The conductance in this case is
\begin{equation}
\label{eq:limits2}
  \begin{split}
  g &
  \approx
  \int_0^1 \form{u} \frac{
      1- u^2 
    }{
      1- u^2 \cos^2{\left(k_F L \xi \right)} 
    } \\
    &= 
    \frac{
      |\cos{(k_F L \xi)}| 
      -  \sin^2{(k_F L \xi)}\arctanh{( |\cos{(k_F L \xi)}| )}
    }{
      |\cos{(k_F L \xi)} |^3
    },
  \end{split}
\end{equation}
which oscillates between the values $2/3$ and $1$ with period $\pi/k_F L$ as
a function of $\xi$.

\subsubsection{Evanescent modes}

When $|E_F - U| \ll |E_F|$, the evanescent modes dominate the transport so we
neglect the contribution from the propagating modes. Using that $|\xi|
\leq u$ for evanescent modes, and that $|\xi| \ll 1$, we find
\begin{equation}
  \label{eq:limits4}
  \begin{split}
    T(u)  &
    \approx
    \frac{
      1
    }{
      \cosh^2{\left(k_F L u\right)}
    },
  \end{split}
\end{equation}
which is valid for $k_F L \gg 1$. The conductance then becomes
\begin{equation}
  \label{eq:limits5}
  \begin{split}
  g &
  \approx \int_0^1 \frac{ 1 }{ \cosh^2{\left(k_F L u\right)} } 
  \approx \frac{1}{k_F L}.
  \end{split}
\end{equation}

\section{Dip of the tunneling conductance around the Dirac point}
\label{sec:length-depend-struct}

We study how the width of the dip in the conductance around the Dirac
point scales with barrier length $L$. The contributions from both
evanescent and from propagating modes must be considered when $|\xi| <
1$.

The conductance due to propagating modes can be written as  
\begin{equation}
  g_\text{prop} =  f(\xi) |\xi|
\end{equation}
where
\begin{equation}
  f(\xi) =
  \int_0^1 \form{v}
  \frac{
    1
  }{
    1
    +
    \frac{
      v^2 \left( 1-\xi \right)^2
    }{
      \left(1-v^2\right)
      \left(1-\xi^2 v^2 \right)
    }
    \sin^2{\left(k_F L \xi \sqrt{1-v^2} \right)}
  }.
\end{equation}
 When $k_F L \gg 1$, $f(\xi)$ is
well approximated by a linear curve $a_1 + a_2 \xi$ for all $|\xi| <
1$. The function $f(\xi)$ deviates from linearity in an oscillatory
fashion in a small region around $|\xi| = 0$, but 
$f(\xi)$ is allways of order unity.
For
$k_F L \gg 1$, the conductance due to propagating waves in the region
$|\xi| < 1$ can therefore be approximated by
\begin{equation}
  \label{eq:c3}
  g_\text{prop} \approx  \left( a_1 + a_2 \xi \right) |\xi|,
\end{equation}
where the value of the constants $a_1 = 0.79$ and $a_2 = 0.21$ depend
weakly on $k_F L$ when $k_F L \gg 1$ and are found by fitting
Eq.~\eqref{eq:c3} to numerical calculations.

We have not been able to obtain an analytical expression for the
contribution due to evanescent modes. However, we note that the
contribution from $T(u)$ in Eq.~\eqref{eq:transm-prob-angle} to
evanescent modes can be well approximated by a decaying exponential
function. We have fitted our numerical calculations of $
g_\text{evan}=\int_{\abs{\xi}}^{1} \form{u} T(u)$ to an exponentially
decaying function of $|\xi|$:
\begin{equation}
  g_\text{evan} \sim A e^{-B |\xi|}.
\end{equation}
The constant $A$ is found to be $1/k_F L$ by letting $\xi \to 0$ and
comparing with Eq.~\eqref{eq:GUequalsE}. Numerical evidence suggest
that $B = C k_F L$, with $C$ of order unity.

We define the width $w$ of the dip in the conductance at the Dirac
point as $w = 2 |\xi_c|$, where $|\xi_c|$ is the value of $|\xi|$ for
which $g_\text{prop}(\xi_c) = g_\text{evan}(\xi_c)$. Taking advantage
of the fact that $g_\text{evan}$ decays rapidly away from $\abs{\xi} =
0$, we ignore the second order term in the expression for
$g_\text{prop}$ for the purpose of estimating the width $w$. We find
that
\begin{equation}
  w = 2 |\xi_c| 
  \approx \frac{1}{k_F L},
\end{equation}
using that $2 \mathcal{W}(1/a_1) \approx 1$, where $\mathcal{W}$ is
the Lambert W-function.

\bibliography{spinsplitting_3}

\end{document}